# The Complexity of Symmetry Breaking in Massive Graphs

## Christian Konrad
Department of Computer Science, University of Bristol
christian.konrad[at]bristol.ac.uk

## Sriram V. Pemmaraju
Department of Computer Science, The University of Iowa
sriram-pemmaraju[at]uiowa.edu

## Talal Riaz
Department of Computer Science, The University of Iowa
sriram-pemmaraju[at]uiowa.edu

## Peter Robinson
Department of Computer Science, City University of Hong Kong
peter.robinson[at]cityu.edu.hk

## Abstract

The goal of this paper is to understand the complexity of symmetry breaking problems, specifically *maximal independent set (MIS)* and the closely related $\beta$-*ruling set* problem, in two computational models suited for large-scale graph processing, namely the $k$-machine model and the graph streaming model. We present a number of results. For MIS in the $k$-machine model, we improve the $\tilde{O}(m/k^2 + \Delta/k)$-round upper bound of Klauck et al. (SODA 2015) by presenting an $\tilde{O}(m/k^2)$-round algorithm. We also present an $\tilde{\Omega}(n/k^2)$ round lower bound for MIS, the first lower bound for a symmetry breaking problem in the $k$-machine model. For $\beta$-ruling sets, we use *hierarchical sampling* to obtain more efficient algorithms in the $k$-machine model and also in the graph streaming model. More specifically, we obtain a $k$-machine algorithm that runs in $\tilde{O}(\beta n \Delta^{1/\beta}/k^2)$ rounds and, by using a similar hierarchical sampling technique, we obtain one-pass algorithms for both insertion-only and insertion-deletion streams that use $O(\beta \cdot n^{1+1/2^{\beta-1}})$ space. The latter result establishes a clear separation between MIS, which is known to require $\Omega(n^2)$ space (Cormode et al., ICALP 2019), and $\beta$-ruling sets, even for $\beta = 2$. Finally, we present an even faster 2-ruling set algorithm in the $k$-machine model, one that runs in $\tilde{O}(n/k^{2-\epsilon} + k^{1-\epsilon})$ rounds for any $\epsilon$, $0 \le \epsilon \le 1$. For a wide range of values of $k$ this round complexity simplifies to $\tilde{O}(n/k^2)$ rounds, which we conjecture is optimal.

Our results use a variety of techniques. For our upper bounds, we prove and use simulation theorems for beeping algorithms, hierarchical sampling, and $L_0$-sampling, whereas for our lower bounds we use information-theoretic arguments and reductions to 2-party communication complexity problems.

# 1 Introduction

The dramatic growth in the size of graphs that need to be algorithmically processed has led to exciting research in large-scale distributed and streaming graph algorithms. Specifically, there has been a flurry of research on graph algorithms and lower bounds in models of large-scale distributed computation such as the *MapReduce* model [24], the *massive parallel computation (MPC)* model [40], and the *k-machine* model [26]. Simultaneously, a lot of progress has been made on designing low-memory graph algorithms and proving memory lower bounds in different data streaming models [31]. The goal of this paper is to understand the complexity of *symmetry breaking* problems, specifically *maximal independent set (MIS)* and the closely related *ruling sets* problem, in the $k$-machine and streaming models. The MIS problem is a fundamental building block in distributed and parallel computing and efficient distributed algorithms for MIS are the basis for efficient distributed algorithms for problems such as *minimum dominating set* and *facility location*.[1] A $\beta$-*ruling set* of a graph $G = (V, E)$, for integer $\beta \geq 1$, is an independent set $I \subseteq V$ such that every node in $V$ is at most $\beta$ hops from some node in $I$. An MIS is just a 1-ruling set and $\beta$-ruling sets for larger $\beta$ are natural relaxations of an MIS.

There is a rich interplay between techniques in distributed algorithms and those in streaming algorithms. On the algorithmic side, $L_0$-sampling [22] and the related linear graph sketches [3], which were first developed in the context of insertion-deletion streams have also been used for optimal algorithms for connectivity and MST in the distributed CONGESTEDCLIQUE [23] and $k$-machine models [26, 35]. On the lower bound side, there are numerous examples of reductions to 2-party or multiparty communication complexity problems being used to derive lower bounds for both distributed computing and streaming problems. In this paper, *hierarchical sampling* is the common technical thread that connects our $k$-machine results and streaming results.

**The $k$-machine model**

The $k$-machine model was introduced by Klauck et al. [26] as an abstraction of the computation performed by large-scale graph processing systems such as Pregel [30] and Giraph (see http://giraph.apache.org/, [11]). This model assumes $k$ machines $m_1, m_2, \ldots, m_k$ connected by a clique communication network. Computation and communication proceed in fault-free, synchronous rounds via message passing, as in standard models of distributed computation such as CONGEST [38]. The typical assumption regarding bandwidth constraints in the $k$-machine model is that in each round, each communication link can carry a message of size $O(\text{poly}(\log n))$ bits. The input consists of a massive $n$-vertex graph, with $n \gg k$. The graph is assumed to be distributed randomly in a vertex-centric fashion, i.e., each vertex and all incident edges are assigned to a machine picked uniformly at random from among the $k$ machines. Thus each machine hosts $\tilde{O}(n/k)$ vertices *with high probability (whp)*[2]. Furthermore, a machine $m_i$ that hosts a vertex $v$ also knows not just the neighbors of $v$, but also the machines that host these neighbors. This assumption about initial knowledge is sometimes referred to as the KT1 ("(K)nowledge (T)ill Radius 1") assumption [6].

In the paper by Klauck et al. [26] and in subsequent works [7, 20, 35, 37], upper and lower bounds for several important graph problems such as *connectivity*, *minimum spanning tree (MST)*, *page rank*, *triangle enumeration*, etc., are shown. For example, an $\tilde{\Omega}(n/k^2)$ round lower bound on connectivity is shown in [26] and a tight (within logarithmic factors) upper bound of $\tilde{O}(n/k^2)$ is shown in [35]. While we have a good understanding of connectivity and related "global" problems in the $k$-machine model, this understanding does not extend to MIS and symmetry breaking problems such as ruling sets. For example, [26] mention an $\tilde{O}(\min\{\frac{n}{k}, \frac{m}{k^2} + \frac{\Delta}{k}\})$-round MIS algorithm in the $k$-machine model that is obtained from a direct simulation of Luby's MIS algorithm [29, 4]. On the other hand, no lower bounds are known for MIS or any related symmetry breaking problems such as $\beta$-ruling sets in the $k$-machine model. In this paper, we shed some light on the complexity of symmetry breaking in the $k$-machine model; our specific results are described in more detail below.

---

[1] The citation for the 2016 Dijkstra prize in Distributed Computing calls MIS the "crown jewel of distributed symmetry breaking problems."

[2] We use $\tilde{O}(f(n))$ and $\tilde{\Omega}(f(n))$ notation to hide polylogarithmic factors. $\tilde{O}(f(n))$ is short for $O(f(n) \log^c n)$ for some constant $c$ and $\tilde{\Omega}(f(n))$ is short for $\Omega(f(n)/\log^c n)$ for some constant $c$. The phrase "with high probability" refers to probability at least $1 - 1/n$.

**Data Streaming Models**

Data streaming algorithms are motivated by the fact that modern data sets are too large to fit into a computer's random access memory. A streaming algorithm processes its input sequentially item by item in one or few passes while maintaining a random access memory of sublinear size in the input [32]. Graph problems have been studied in the streaming model for roughly 20 years [19] (see also [31] for a more recent survey). Given an $n$-node graph $G$, a streaming algorithm processing $G$ makes one or few passes over the edges of $G$. A priori no assumption is made regarding the order in which the edges "arrive". We will also consider graph streams that consist of both edge insertions and deletions (also known as *dynamic* or *turnstile* streams), where an edge can only be deleted if it has previously been inserted. This model has first been investigated by Ahn et al. [3] and has since been the focus of active research.

It is known that space $\Omega(n^2)$ space is necessary for one pass streaming algorithms [12, 5] that solve MIS, i.e., the trivial algorithm that stores all edges and computes a maximal independent set in the post-processing stage is optimal. However, nothing is known about ruling sets in the streaming model. Using a hierarchical sampling approach we show there is a $\beta$-ruling set algorithm using $o(n^2)$ space, showing a clear separation between the space complexity of $\beta$-ruling sets for $\beta = 1$ and $\beta = 2$ in the one pass streaming setting.

## 1.1 Main Contributions

The contributions of this paper can be organized into three categories as follows.

**MIS bounds.** We present an $\tilde{O}\left(\min\{\frac{n}{k}, \frac{m}{k^2}\}\right)$-round MIS algorithm in the $k$-machine model for graphs with $m$ edges, improving on the $\tilde{O}\left(\min\{\frac{n}{k}, \frac{m}{k^2} + \frac{\Delta}{k}\}\right)$-round MIS algorithm of Klauck et al. [26]. This result follows from a more general result, namely a simulation theorem that shows that any *beeping* algorithm with message complexity $msg$, running in $T$ rounds can be simulated in the $k$-machine model in $\tilde{O}(msg/k^2 + T)$ rounds. Beeping algorithms [1, 18, 25, 41] use extremely simple communication – just *beeps* – and node actions in a round only depend on whether a node has heard a beep (or not) in this round. Our result illustrates a general theme: algorithms in standard models of distributed computation can be automatically translated into efficient algorithms in models of large-scale distributed computing if they (i) use simple communication and (ii) if they have low round complexity *and* message complexity.

We also present an $\tilde{\Omega}(n/k^2)$ lower bound for MIS, the first non-trivial lower bound for a symmetry breaking problem in the $k$-machine model. Our proof starts by showing that in the 2-party communication complexity setting, there is a $O(1)$-sized graph gadget for which Alice and Bob need to communicate $\Omega(1)$ bits to find an MIS. We use an information-theoretic argument to show this and then using a direct sum type argument, we amplify this result to show an $\Omega(n)$ lower bound on the communication complexity of MIS. We then reduce the $k$-machine MIS problem to the 2-party MIS problem to obtain the result, which holds even for randomized algorithms with a constant error probability. It is worth noting that this approach does not yield a 2-ruling set lower bound since 2-ruling sets can be computed in the 2-party setting without *any* communication!

**Hierarchical sampling for ruling set upper bounds.** We use hierarchical sampling to obtain a $k$-machine $\beta$-ruling set algorithm, for $\beta > 1$, that is faster than the fastest known MIS algorithm. A similar hierarchical sampling approach also leads to one-pass $\beta$-ruling set algorithms in both the insertion-only and the insertion-deletion edge-streaming models that use strictly subquadratic space. Specifically, for the $\beta$-ruling set problem, we present an $\tilde{O}(\beta \cdot n\Delta^{1/\beta}/k^2)$-round algorithm in the $k$-machine model and one-pass streaming algorithms using space $\tilde{O}(\beta \cdot n^{1+\frac{1}{2^{\beta-1}}})$. Our $k$-machine $\beta$-ruling set algorithm is faster than the fastest known MIS algorithm, even for $\beta = 2$. But this result does not imply a separation between MIS and 2-ruling set in the $k$-machine model since we only know an $\tilde{\Omega}(n/k^2)$ lower bound for MIS. However, the streaming algorithm we present implies a clear separation between MIS and 2-ruling sets in this model due to the $\Omega(n^2)$ space lower bound for MIS [12, 5]. For insertion and deletion streams, we use $L_0$-*sampling* [22] as the basic building block of our algorithm.

**Faster $k$-machine 2-ruling set algorithm.** For the special case of 2-ruling sets, we present an even faster algorithm, one that runs in $\tilde{O}(n/k^{2-\epsilon} + k^{1-\epsilon})$ rounds for any $\epsilon$, $0 \le \epsilon \le 1$. For $\epsilon = 0$, this yields an $\tilde{O}(n/k^2 + k)$-round algorithm, which simplifies to $\tilde{O}(n/k^2)$ rounds for $k \le n^{1/3}$. We conjecture that $\tilde{\Omega}(n/k^2)$ is a lower bound for 2-ruling sets in the $k$-machine model, and proving this would show that the above-mentioned upper bound is tight. This algorithm uses a combination of greedy-style sequential processing technique that is tailored to the $k$-machine model, and a beeping version of the low message complexity 2-ruling set algorithm of [34, 33] originally designed for the CONGEST model.

## 1.2 Related Work

The fastest MIS algorithm in the classical LOCAL and CONGEST models of distributed computin is still the three-decade old algorithm due to Luby [29] and independently due to Alon, Babai, and Itai [4]. This algorithm runs in $O(\log n)$ rounds and closing the gap between this upper bound and the $\Omega\left(\min\left\{\sqrt{\frac{\log n}{\log \log n}}, \frac{\log \Delta}{\log \log \Delta}\right\}\right)$ lower of Kuhn, Moscibroda, and Wattenhofer [28] is a major open question in this area. Assuming a bounded maximum degree, faster MIS algorithms have been very recently designed for both the LOCAL model [9, 14] and the CONGEST model [15]. $\beta$-ruling sets have also recently garnered interest in the LOCAL and CONGEST models [10, 27, 9, 10, 14, 15] and the fastest 2-ruling set algorithm in the LOCAL model breaks the Kuhn-Moscibroda-Wattenhofer lower bound and runs faster than any MIS algorithm can.

Research on algorithms and lower bounds in the $k$-machine model has been mentioned earlier in the introduction. The *massive parallel computation (MPC)* model is related to the $k$-machine model, but there are important differences in local memory and bandwidth constraints between the models. The study of classical symmetry breaking problems, especially MIS, in the MPC model is a very active area of current research [16, 17].

Many of the classic symmetry breaking problems in distributed computing have been studied in the streaming model. As mentioned earlier, there is a space $\Omega(n^2)$ lower bound for MIS for one pass algorithms [12, 5]. If multiple passes are granted, it is possible to use the correlation clustering algorithm of [2] to compute an MIS in $p$ passes using space $\tilde{O}(n^{1+\frac{1}{2^{p-1}}})$. A *maximal matching* can easily be maintained in the streaming model with space $\tilde{O}(n)$, by running the GREEDY matching algorithm. Similar to the distributed setting where computing a $(\Delta + 1)$-coloring is easier than computing an MIS, in a recent breakthrough, Assadi et al. [5] gave a one-pass streaming algorithm with space $\tilde{O}(n)$ for $(\Delta + 1)$-coloring, even in insertion-deletion streams.

**Remark:** Proofs that are omitted from Sections 2.2 and 3.1. are included in the appendix.

## 2　Upper and Lower Bounds for MIS

### 2.1　An $\tilde{O}(m/k^2)$ upper bound for $k$-machine MIS

This section presents an $\tilde{O}(\frac{m}{k^2})$-round MIS algorithm in the $k$-machine model, improving on the current fastest MIS algorithm due to Klauck et al. [26] that runs in $O(\frac{m}{k^2} + \frac{\Delta}{k})$ rounds[3] The Klauck et al. MIS algorithm is simply obtained by simulating Luby's MIS algorithm in the $k$-machine model. Here we show a general result first, that *beeping model* algorithms [1, 18, 25, 41] can be efficiently simulated in the $k$-machine model and then apply this result to the $O(\log n)$-round beeping model MIS algorithm of Jeavons et al. [21].

The *beeping model* assumes a network of nodes that synchronously communicate, but only in *beeps*. A node in this model can distinguish between two situations in a round: (i) no neighbor has beeped versus (ii) at least one neighbor has beeped. The beeping model is motivated by communication in wireless networks [18, 25] and also in biological processes that solve complex problems using very simple messages e.g., neural precursor selection in the *Drosophilla* fly [1]. In both of these applications, it is found that despite the simplicity of communication, beeping model algorithms are quite powerful. Our motivation for simulating beeping algorithms in the $k$-machine model is similar; since a beeping algorithm has simple communication, it is easy to simulate it efficiently in the $k$-machine model, yet for some problems (e.g., MIS) beeping algorithms seem as powerful as algorithms that use more complex communication schemes.

To state the efficiency of our simulation, we need to define the message complexity of a beeping algorithm. Viewing each beep as a broadcast, we assume that a node $v$ sends $degree(v)$ messages whenever it beeps. We define *message complexity*, $msg(A)$, of an algorithm $A$ in the beeping model as the total number of messages sent during the course of the algorithm. The simulation itself is simple. Each machine performs local computations on behalf of all nodes it hosts and then sends and receives messages (beeps) on behalf of these nodes. The simulation can be done efficiently because each machine can aggregate beeps in two ways. First, if a node $v$ hosted by machine $M$ has several neighbors hosted by machine $M'$, then $M$ needs to send just one beep on $v$'s behalf to its neighbors in $M'$. This aggregation works for any *broadcast* algorithm and it is exploited in the Conversion Theorem in [26]. Additional aggregation is possible because the algorithm is in the beeping model. Specifically, if $M$ hosts several nodes $u_1, u_2, \ldots, u_p$ that have a common neighbor $v$ hosted by $M'$, then $M$ can send just one beep on behalf of all of $u_1, u_2, \ldots, u_p$ to $v$ in $M'$.

▶ **Theorem 1.** *A beeping algorithm $A$ that runs in $T$ rounds can be implemented in the $k$-machine model in $\tilde{O}(\frac{msg(A)}{k^2} + T)$ rounds.*

**Proof.** Let $a_t$ denote the message complexity of algorithm $A$ in round $t$. A node that beeps in round $t$ is said to be *active* in round $t$. (Note that $a_t$ is the sum of the degrees of nodes that are active in round $t$.) Partition the active nodes in round $t$ by their degree into $O(\log \Delta)$ degree classes; $[1, 2), [2, 4), [4, 8), \ldots, [\Delta/2, \Delta), [\Delta, 2\Delta)$. Consider a degree class $[d, 2d)$ and let $n_d$ denote the number of active nodes in round $t$ in this class.

---

[3]Klauck et al. also point out that there is simple $\tilde{O}(n/k)$-round MIS $k$-machine algorithm. This allows Klauck et al. to state the running time as $\tilde{O}\left(\min\left\{\frac{n}{k}, \frac{m}{k^2} + \frac{\Delta}{k}\right\}\right)$. Our result improves this to $\tilde{O}\left(\min\left\{\frac{n}{k}, \frac{m}{k^2}\right\}\right)$.

▷ **Claim 2.** A machine sends $\tilde{O}(\frac{a_t}{k} + k)$ messages *whp* for active nodes in degree class $[d, 2d)$ in the simulation of round $t$.

**Proof. Case 1**: $n_d = \Omega(k \log n)$. Since each node in this class has degree at least $d$, the number of active nodes in this degree class is $\leq \frac{a_t}{d}$. Since nodes in the $k$-machine model are distributed uniformly at random among the machines, the expected number of active nodes hosted at a machine in this degree class is $\frac{a_t}{dk}$. Since $n_d = \Omega(k \log n)$, we have that $\frac{a_t}{dk} = \Omega(\log n)$. Thus, we can use a Chernoff bound to show that the number of active nodes in this degree class hosted at any machine is $O(\frac{a_t}{dk})$ *whp*. Since each node in this degree class sends at most $2d$ messages in round $t$, a machine needs to send at most $\tilde{O}(\frac{a_t}{dk} \cdot 2d) = \tilde{O}(\frac{a_t}{k})$ messages for this degree class *whp*.

**Case 2**: $n_d = O(k \log n)$. Using a Chernoff bound, we see that each machine has $O(\log n)$ active nodes from this degree class *whp*. Since $A$ is a beeping algorithm, we need to send at most $k$ beeps (one for each machine) for any node in the simulation of round $t$.

The claim follows from combining the round complexity from the two cases. ◀

Since $[d, 2d)$ is any arbitrary degree class and there are a total of $O(\log \Delta)$ degree classes, each machine sends a total of $\tilde{O}(\frac{a_t}{k} + k)$ messages to simulate round $t$. We can repeat the above argument by categorizing nodes by round $t$ *in-degree*, i.e., the number of neighbors of a node that have beeped in round $t$. We then use the fact that in the beeping model messages incoming to a node can also be aggregated and thus a machine needs to receive at most $k$ messages for any node it hosts. We conclude that a machine *receives* $\tilde{O}(\frac{a_t}{k} + k)$ messages *whp* in the simulation of round $t$.

Next, we argue that all the machines can send and receive all these messages in the $k$-machine model in $\tilde{O}(\frac{a_t}{k^2} + 1)$ rounds. For this we appeal to the following claim on the round complexity of a simple, randomized routing scheme (shown in Algorithm 1). In this scheme, each machine randomly selects a batch of size $k$ messages and then distributes these to the $k$ machines randomly as intermediate destinations. Then the intermediate nodes deterministically send these messages on to their final destinations.

▷ **Claim 3.** Suppose that for some positive integer $X$, each machine has at most $X$ messages to send and each machine is required to receive at most $X$ messages. Then Algorithm 1 delivers all of these messages in $\tilde{O}(X/k)$ rounds.

**Proof.** The number of marked messages at a machine is $\tilde{O}(k)$ *whp*. Therefore, Step 2 takes $\tilde{O}(1)$ rounds *whp*. We must now show that each machine hosts at most $\tilde{O}(1)$ messages intended for a particular destination at the beginning of Step 3.

Consider machines $m_i, m_j$. Again, the number of messages intended for $m_i$ marked in Step 1 is also $\tilde{O}(k)$ *whp*. Since a message intended for $m_i$ ends up at $m_j$ with probability at most $1/k$, the expected number of messages intended for $m_i$ that end up at $m_j$ at the end of Step 2 is $\tilde{O}(1)$.

Consider a message $msg$ that is intended for a machine $m_j$. Let $E_{msg,i,j}$ denote the event that message $msg$ ends up at machine $m_j$ at the end of Step 2. Let $X_{msg,i,j}$ denote the indicator random variable is 1 iff event $E_{msg,i,j}$ occurs, otherwise it is 0. Now, consider another message $msg'$. If $msg$ and $msg'$ are hosted at different machines at the start of the algorithm, then the random variables $X_{msg,i,j}$ and $X_{msg',i,j}$ are independent. On the other hand, if $msg$ and $msg'$ were hosted at the same machine at the start of the algorithm, then the random variables $X_{msg,i,j}$ and $X_{msg',i,j}$ are negatively correlated. Then, using a Chernoff bound for negatively correlated random variables, we see that the number of messages intended for machine $m_i$ that end up at machine $m_j$ is $\tilde{O}(1)$ *whp* [13]. Finally, using a union bound on the total number of $i, j$ pairs, each machine hosts $\tilde{O}(1)$ messages destined for every other machine at the end of Step 2. This means that Step 3 can be completed in $\tilde{O}(1)$ rounds, and our claim about Algorithm 1 holds. ◀

We use Algorithm 1 to route all messages in the simulation of round $t$ $\tilde{O}(\frac{a_t}{k^2} + 1)$ rounds. Since $t$ is any arbitrary round, we can simulate all $T$ rounds of $A$ in the $k$-machine model in $\sum_{1 \leq t \leq T} \tilde{O}(\frac{a_t}{k^2} + 1) = \tilde{O}(\frac{msg(A)}{k^2} + T)$ rounds. This completes the proof of the theorem. ◀

---

**Algorithm 1:** RANDOMIZEDROUTING

**1** Each unsent message that a machine hosts is marked with probability $\min(\frac{k}{Y}, 1)$, where $Y$ is the current number of unsent messages the machine holds.

**2** Each machine distributes marked messages it holds to the $k$ machines by picking a random permutation of these messages, and sending the $i^{th}$ message in the permutation to machine $m_{i^*}$ where $i^* := i \mod k$. If a machine holds several messages intended for a particular destination, then it sends these messages one-by-one.

**3** Each machine deterministically sends marked messages received in each round to their final destination. If a machine holds several messages intended for a particular destination, it will just send these messages one-by-one.

---

The following result is immediate by applying the simulation result above to the $O(\log n)$-round beeping model MIS algorithm of Jeavons et al. [21].

▶ **Theorem 4.** *MIS can be computed in $\tilde{O}(\frac{m}{k^2})$ rounds in the $k$-machine model, where $m$ is the number of edges in the input graph.*

## 2.2 An $\tilde{\Omega}(n/k^2)$ lower bound for $k$-machine MIS

In this section, we show an $\tilde{\Omega}(\frac{n}{k^2})$ lower bound for MIS. While numerous lower bounds have been shown for the $k$-machine model for problems such as pagerank approximation, triangle enumeration, and graph connectivity (see [35, 37, 26]), these techniques cannot be applied directly to our setting. The reason for this is that the proof technique in previous work heavily relies on the fact that the input graph determines the *unique* correct solution, whereas, there are many feasible maximal independent sets for a given input graph.

In our proof we proceed as follows: We start out by considering the problem in the 2-party communication model of [39]. In particular, in Section 2.2.1 we first prove an $\Omega(1)$ communication complexity lower bound for solving MIS on a constant size gadget, which we subsequently extend to a lower bound for solving $\Theta(n)$ independent copies of the gadget. In Section 2.2.3, we describe how to extend this result to the $k$-machine model.

### 2.2.1 A 2-party MIS Lower Bound For a Single Gadget

▶ **Theorem 5.** *The two party communication complexity of MIS on constant-size graphs is $\Omega(1)$.*

In the remainder of this section we prove Theorem 5. We define a 7-digit vector $s = s_1 s_2 \ldots s_7$ as *valid* if each $s_i$ is in the range $[1, 7]$ and for exactly two of its digits $s_i \neq i$. Moreover, it must be that if $s_i \neq i$ and $s_j \neq j$, then $s_i = j$ and $s_j = i$. Suppose that Alice and Bob receive inputs $X$ and $Y$ chosen uniformly at random from all valid 7 digit vectors. Since there are 21 such valid vectors, and $X$ and $Y$ are chosen uniformly at random from all valid vectors, $H[X] = H[Y] = \log_2 21$.

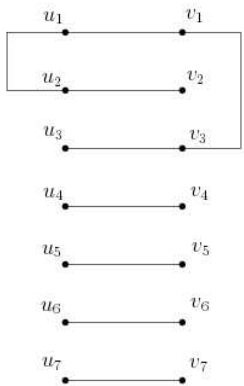

**Figure 1** The lower bound gadget $g$ with $X = 2134567$ and $Y = 3214567$

We will show that Alice and Bob can construct a gadget $g$ based on the inputs $X$ and $Y$ such that for any MIS $I$ of $g$, either the conditional mutual information, $\mathbf{I}[Y : I \mid X]$ or $\mathbf{I}[X : I \mid Y]$ is $\Omega(1)$. As it is well known that mutual information lower bounds communication complexity [8], this immediately implies the lower bound claimed in Theorem 5.

**The lower bound gadget**

The gadget $g$ consists of 14 nodes, $V_A = \{u_1, \ldots, u_7\}$ and $V_B = \{v_1, \ldots, v_7\}$. Conceptually, the $u$ nodes are hosted at Alice and the $v$ nodes are hosted at Bob. The gadget contains 7 edges $(u_i, v_i)$ for $i \in [1, 7]$. Additionally, the gadget also contains two special edges that are determined by the inputs $X$ and $Y$. Based on the input $X = x_1 \ldots x_7$, Alice will add a single edge between a pair of $u$ nodes. Specifically, for the two indices $i$ and $j$ ($i \neq j$) in its input vector where $x_i = j$ and $x_j = i$, Alice adds the edge $(u_i, u_j)$. Thus, each valid vector $X$ corresponds to a unique edge between the $u$ nodes. Similarly, Bob will add one edge to its nodes based on its input $Y$. We call these two edges *special edges*. Note that except for the two special edges, the topology of the gadget is independent of the inputs $X$ and $Y$.

The following lemma suffices to prove Theorem 5.

▶ **Lemma 6.** *Let $I_A$ resp. $I_B$ denote the vertices in the MIS output by Alice resp. Bob. Either $\mathbf{I}[X : I_B \mid Y] = \Omega(1)$ or $\mathbf{I}[Y : I_A \mid X] = \Omega(1)$.*

### 2.2.2 A 2-Party Lower Bound for Multiple Gadgets

Recalling that any pair of valid bit vectors $(X, Y)$ uniquely defines the topology of a gadget, we call $(X, Y)$ the *value of the gadget* and, to simplify the notation, we also use $(X, Y)$ to refer to the gadget itself.

▶ **Theorem 7.** *The two party communication complexity of MIS on graphs with $O(n)$ nodes and edges is $\Omega(n)$.*

In the rest of this subsection, we prove Theorem 7. Consider again the two party communication complexity model where Alice receives input $X$ and Bob receives input $Y$. We now consider $X$ and $Y$ to be vectors of length $n/2$ and we conceptually think of such a vector as the concatenation of $n/14$ 7-digit vectors as defined in Section 2.2.1. We say that an $(n/2)$-length digit vector $S = s_{1,1} \ldots s_{1,7} s_{2,1} \ldots s_{2,7} \ldots s_{n/14,7}$ is *valid*, if $(s_{i,1} \ldots s_{i,7})$ forms a *valid* 7-digit vector, for all $i \in [1, n/14]$. Let $X$ and $Y$ be two $n/2$ digit vectors chosen uniformly at random from all valid $(n/2)$-length digit vectors. According to the inputs $X$ and $Y$, Alice and Bob will construct the lower bound graph $G_L$, having the property that, for any MIS $I$ of $G_L$, either the mutual information between Alice's MIS output and Bob's input or between Bob's MIS output and Alice's input is $\Omega(n)$.

We now describe how to construct the lower bound graph. $G_L$ contains $n$ nodes partitioned into sets $V_A := \{u_1, \ldots, u_{n/2}\}$ and $V_B := \{v_1, \ldots, v_{n/2}\}$. All the nodes in $V_A$ are hosted at Alice and all the nodes in $V_B$ are hosted at Bob. The edges of $G_L$ induce $n/14$ gadgets $\{g_1, \ldots, g_{\frac{n}{14}}\}$ where gadget $g_i$ contains nodes $u_{7i+1}, \ldots, u_{7i+7}$ and $v_{7i+1}, \ldots, v_{7i+7}$. The value of gadget $g_i$ is $(x_{i,1} \ldots x_{i,7}, y_{i,1} \ldots y_{i,7})$. For example, gadget $g_1$ contains nodes $u_1, \ldots, u_7$ and $v_1, \ldots, v_7$, and has value $(x_{1,1} \ldots x_{1,7}, y_{1,1} \ldots y_{1,7})$. Notice that the topology of $G_L$ depends only on the inputs $X$ and $Y$.

Theorem 7 follows immediately from the following lemma.

▶ **Lemma 8.** *Either* $\mathbf{I}[X : I_B \mid Y] = \Omega(n)$ *or* $\mathbf{I}[Y : I_A \mid X] = \Omega(n)$.

### 2.2.3 Extension to the $k$-machine model

We are now ready to extend our results from the 2-party communication setting to the $k$-machine model. Specifically, we want to show a lower bound for $\varepsilon$ error (possibly randomized) algorithms i.e., algorithms which, over all graph partitions (and random coin tosses), outputs an MIS with probability at least $(1 - \varepsilon)$, and always terminates in $T$ rounds.

▶ **Theorem 9.** *For a constant $\varepsilon > 0$, any $\varepsilon$-error (possibly randomized) MIS algorithm in the $k$-machine model has round complexity at least $\tilde{\Omega}(\frac{n}{k^2})$.*

## 3 Ruling sets via hierarchical sampling

### 3.1 An Algorithm for the $k$-machine Model

In Algorithm 2 we use *hierarchical sampling* to compute a $\beta$-ruling set of an $n$-vertex graph with maximum degree $\Delta$ in $\tilde{O}(n\Delta^{1/\beta}/k^2)$ rounds. A hierarchical sampling approach has also been used in [10] to compute $\beta$-ruling sets and combined with the MIS algorithms of Ghaffari [14, 15], yields the fastest $\beta$-ruling set algorithms in the LOCAL and CONGEST models. But there are key differences between the LOCAL/CONGEST algorithms and our $k$-machine algorithm because the bandwidth constraints of the $k$-machine model are quite stringent (for e.g., it seems difficult for nodes that are deactivated to efficiently inform their neighbors of this fact in the $k$-machine model).

In each iteration $i$, $2 \leq i \leq \beta$, in Algorithm 2, we independently sample active nodes with probability $\Theta(\log n/\Delta^{1-(i-1)/\beta})$ (Step (3)). In each iteration, we get a sampled graph that is communicated across the $k$ machines (Step (5)) and then we compute an MIS on this sampled graph (Step (6)). Note that in order to communicate the sampled graph, each sampled node needs to communicate with *all* neighbors and not just sampled neighbors because a priori a node does not know which neighbors have been sampled. Thus for the communication step to be efficient, i.e., complete in $\tilde{O}(n\Delta^{1/\beta}/k^2)$ rounds, as shown in Lemma 12, nodes participating the sampling step need to have relatively low degree. To ensure this high degree nodes need to be deactivated in each iteration. A node that has more than $\Delta^{1-(i-1)/\beta}$ neighbors participating in the sampling step in Iteration $i$ is guaranteed (whp) to have a sampled neighbor and such a node will deactivate itself if it is not marked. But, a node may have high degree but with only few neighbors participating in the sampling step. A node $v$ of this type is guaranteed to have neighbors that were deactivated in previous iterations. By inductively assuming that a node deactivated in an earlier round is not too far away from some node that has joined the ruling set, we get that node $v$ itself is at most one extra hop away from the ruling set and can be deactivated (as in Step (7)).

---

**Algorithm 2:** RAND $\beta$-RULING SET(Graph $G = (V, E)$):

1  $P_1 \leftarrow V$
2  **for** *iteration* $i \leftarrow 2$ *to* $\beta$ **do**
3  $\quad$ Each node in $P_{i-1}$ marks itself with probability $\Theta(\log n/\Delta^{1-(i-1)/\beta})$.
4  $\quad M_i \leftarrow$ nodes marked in the previous step
5  $\quad$ Each node in $M_i$ informs all neighbors that it is marked
6  $\quad I_i \leftarrow MIS(G[M_i])$
7  $\quad$ Each unmarked node that has a neighbor in $M_i$ or has degree $> \Delta^{1-(i-1)/\beta}$ joins the set $T_i$ and is *deactivated*
8  $\quad P_i \leftarrow P_{i-1} \setminus (M_i \cup T_i)$
9  **end**
10 Each node in $P_t$ informs neighbors that it is in $P_t$
11 $I \leftarrow MIS(G[P_t])$
12 **return** $(\cup_j I_j) \cup I$

---

▶ **Lemma 10.** *For $1 \leq i \leq \beta$, the maximum degree of nodes in $P_i$ is at most $\Delta^{1-(i-1)/\beta}$.*

▶ **Lemma 11.** *For $2 \leq i \leq \beta$, the maximum degree of the induced graph $G[M_i]$ is at most $\tilde{O}(\Delta^{1/\beta})$ whp.*

▶ **Lemma 12.** *The communication in Steps (5) and (10) can each be completed in $\tilde{O}(n\Delta^{1/\beta}/k^2)$ rounds whp. The MIS computation in Steps (6) and (11) can each be completed in $\tilde{O}(n\Delta^{1/\beta}/k^2)$ rounds whp.*

▶ **Lemma 13.** *Every node in $V$ is at most $\beta$ hops from some node in $(\cup_{j=2}^{\beta} I_j) \cup I$.*

▶ **Theorem 14.** *For any integer $\beta \geq 1$, a $\beta$-ruling set of an $n$-vertex graph with maximum degree $\Delta$ can be computed in $\tilde{O}(\beta \cdot n \cdot \Delta^{1/\beta}/k^2)$ rounds.*

## 3.2 A One-Pass Edge-Streaming Algorithm

We now use the hierarchical sampling approach to obtain a low-memory algorithm for $\beta$-ruling sets in the edge streaming model, for $\beta > 1$. As mentioned in Section 1, our algorithm stands in contrast to the $\Omega(n^2)$ space lower bound for MIS of [12].

For each $i \in [1, \beta]$, we define

$$q_i := \frac{1}{2^{\beta-i}}. \tag{1}$$

Initially, we subsample a hierarchy of vertex sets $P_1, \ldots, P_\beta$ as follows:
- $P_1 = V(G)$.
- For $i \in [2, \beta]$, and each $u \in P_{i-1}$, we add $u$ to $P_i$ with probability $1/n^{q_{i-1}}$.

We call $\ell(u) = \max\{i \mid u \in P_i\}$ the *level of $u$*, and define the *active degree of $u$* as the node degree of $u$ in the graph induced by the vertices $P_{\ell(u)} \cup \cdots \cup P_\beta$. Note that if $\ell(u) = 1$, then the active degree is simply the node degree.

**Single Pass:** We now describe which edges we store during the single pass of the algorithm. For each $u$, we store the first $\mu_{\ell(u)} := \Theta(n^{q_{\ell(u)}} \log n)$ edges that contribute to $u$'s active degree, i.e., connect $u$ to nodes in $P_{\ell(u)} \cup \cdots \cup P_\beta$; recall that $P_{\ell(u)} \cup \cdots \cup P_\beta \subseteq P_{\ell(u)}$. Observe that (1) implies that we store all incident edges for vertices in $P_\beta$.

Upon storing the $\mu_{\ell(u)}$-th edge for $u$, we mark $u$ as *covered*. While processing the stream, we discard all edges that connect two nodes if they are both marked as covered.

**Post-Processing:** After the pass is completed, we move every $u \notin P_\beta$ that is *not* covered to the set $P_\beta$ and set $\ell(u) = \beta$. As the last step of the algorithm, we compute an MIS $S$ on the graph spanned by the stored edges of nodes in $P_\beta$ (after post-processing) and output $S$ as the result.

▶ **Lemma 15.** *The algorithm outputs a $\beta$-ruling set with high probability.*

**Proof.** We first prove that the resulting output $S$ is indeed an independent set. To this end, we need to show that the following claim holds (after the post-processing step): For every $u, v \in P_\beta$, if there exists $(u, v) \in G$ then also $(u, v) \in E(P_\beta)$, where $E(P_\beta)$ refers to the stored edges incident to nodes in $P_\beta$: Assume towards a contradiction that the claim is false, i.e., the algorithm did not store $(u, v)$. We distinguish 3 cases:
1. If both $u$ and $v$ had level $\beta$ *before* post-processing, then we would have stored $(u, v)$, as we store all incident edges for nodes in $P_\beta$.
2. Suppose only $u$ is moved from its previous level $i$, whereas $v$ was already in $P_\beta$ after the initial sampling. It follows that $u$ was not covered and hence, by definition, the edge $(u, v)$ must have been among the first $\mu_{\ell(u)}$ edges in the stream that were incident to $u$ and that have their other endpoint in $P_i \cup \cdots \cup P_\beta$. By the description of the algorithm, we would have stored $(u, v)$, yielding a contradiction.
3. Finally, suppose $u$ and $v$ were both moved to $P_\beta$ and assume (wlog) that $\ell(u) \leq \ell(v)$ before the post-processing step. By a similar argument as in the previous case, it follows that we would have stored the edge $(u, v)$ as one of the first $\mu_{\ell(u)}$ incident edges of $u$ that point to $P_{\ell(u)} \cup \cdots \cup P_\beta$, again resulting in a contradiction.

Next, we show that the output set $S$ satisfies the distance property of $\beta$-ruling sets, i.e., we argue that every node has distance at most $\beta$ from some node in $S$ with high probability. Recalling that we compute an MIS on the graph induced by $P_\beta$, this clearly holds for any node that has level $\beta$ at this point. Thus, consider a node $u$ with level $\ell(u) = i < \beta$ and assume that $u$ was not moved, i.e., $i$ continues to be the highest level of $u$ after the post-processing step. Since $u \notin P_\beta$, we know that $u$ was covered and hence we stored at least $\mu_i$ many edges incident to $u$ that point to $P_i \cup \cdots \cup P_\beta$. Let $N_i(u)$ denote the corresponding set of neighbors of $u$ for which the algorithm stores edges, i.e., $|N_i(u)| \geq \mu_i = n^{q_i} \log n$. Note that if a neighbor of $u$ in $N_i(u)$ is moved to $P_\beta$ in the post-processing step, this can only reduce the distance of $u$ to a node in the independent set. Therefore, it is sufficient if we show that at least one of $u$'s neighbors in $N_i(u)$ is also part of some level greater than $i$. The probability that none of the $\mu_i$ nodes in $N_i(u)$ is in $P_{i+1} \cup \cdots \cup P_\beta$ is at most

$$\left(1 - \Theta\left(\frac{1}{n^{q_i}}\right)\right)^{\Theta(n^{q_i} \log n)} \leq \frac{1}{n^{\Omega(1)}}.$$

The result follows by taking a union bound over all the nodes. ◀

▶ **Lemma 16.** *The algorithm uses $O\left(\beta \cdot n^{1+1/2^{\beta-1}} \log n\right)$ space with high probability.*

**Insertion-Deletion Streams**

We now describe how to modify the above algorithm to work for insertion-deletion streams. The key observation is that the hierarchical sampling is done completely independently of the input stream and can be done beforehand. The only task remaining while processing the stream is storing a certain number of incident edges to a specific vertex that are also incident to a specific set. In insertion-only streams, this is straightforward. In insertion-deletion streams, we can simply use enough $L_0$ samplers:

Given a stream of edge insertions and deletions, an $L_0$-sampler is able to output a uniform random edge of the input graph (the graph obtained after all insertions and deletions have been applied). This technique can be adapted to most edge sampling tasks, such as sampling a uniform random edge incident to a specific vertex, or, by employing $\Theta(k \log n)$ $L_0$-samplers, sampling $k$ different edges incident to a specific vertex, as it is required in our setting. Jowhari et al [22] showed how to implement an $L_0$-sampler in insertion-deletion streams in small space:

▶ **Theorem 17** ([22]). *There exists an $L_0$ sampler for insertion-deletion streams that uses space $O(\log^2 n \log(1/\delta))$ and succeeds with probability $1 - \delta$.*

For instance, say we want to store $\alpha$ edges incident to a specific vertex $v$. Leveraging Theorem 17 tells us that we only add a polylogarithmic overhead by running $\Theta(\alpha \log n)$ $L_0$ samplers, and we can recover at least $\alpha$ different edges incident to $v$. Together with Lemmas 15 and 16, this implies the following result:

▶ **Theorem 18.** *In both, the insertion-only and the insertion-deletion models, there are randomized one-pass streaming algorithms with space $\tilde{O}(\beta \cdot n^{1 + \frac{1}{2\beta - 1}})$ for computing a $\beta$-ruling set that succeed with high probability.*

## 4 Faster 2-ruling sets in the $k$-machine model

The hierarchical sampling approach from Section 3.1 yields a $k$-machine, 2-ruling set algorithm running in $\tilde{O}(n\Delta^{1/2}/k^2)$ rounds. In this section we use a different approach to obtain a $k$-machine 2-ruling set algorithm that runs in $\tilde{O}(n/k^{2-\epsilon} + k^{1-\epsilon})$ rounds for any $\epsilon$, $0 \le \epsilon \le 1$. Setting $\epsilon = 0$ yields an $\tilde{O}(n/k^2 + k)$-round algorithm; for $k \le n^{1/3}$ this is an $\tilde{O}(n/k^2)$-round algorithm. The optimal value of $\epsilon$, i.e., the value that minimizes the expression $n/k^{2-\epsilon} + k^{1-\epsilon}$, turns out to be $\epsilon = \frac{1}{2}\left(3 - \frac{\log n}{\log k}\right)$. For example, for $k = \sqrt{n}$, $\epsilon = 1/2$ is optimal and the running time simpifies to $\tilde{O}(n^{1/4})$ rounds.

Our algorithm consists of two phases. In the first phase, we perform $\lceil k^\epsilon \rceil$ iterations where we process the input graph in a sequential fashion. We say that a vertex is *active* if its MIS-status is yet undefined; otherwise we say that it is *deactivated*. In iteration $i \ge 1$ of Phase 1, the machine $m_i$ locally computes an MIS $S_i$ on its part of the input and then sends $S_i$ to all other machines using an intermediate routing step. In more detail, after computing the MIS, $m_i$ sends the vertices in $S_i$ in batches of size $k - 1$, by transmitting the vertex IDs of the first $k - 1$ vertices in $S_i$ to the other machines over its $k - 1$ links. The other machines simply relay these messages by broadcast. It is easy to see that all machines know about all nodes in $S_i$ after $m_i$ has sent $O(S_i/k) = \tilde{O}(n/k^2)$ batches. Before proceeding to the next iteration, each machine locally deactivates every vertex that has a neighbor in $S_i$.

The remaining active nodes form the residual graph and machine $m_{i+1}$ operates on its local part of this graph in the next iteration, and so forth. After we have finished all $\lceil k^\epsilon \rceil$ iterations, each machine simply deactivates all of its vertices that are still active and have a neighbor on some machine $m_j$, for $j \in [1, \lceil k^\epsilon \rceil]$. In Lemma 19 below, we show that with high probability only vertices with (initial) degree $\tilde{O}(k^{1-\epsilon})$ remain active after Phase 1. We define $I$ to be the independent set obtained by Phase 1.

For Phase 2, we use the *low message complexity* 2-RULING SET algorithm of Pai et al. [34, 33]. This algorithm runs in the CONGEST model in $O(\Delta \log n)$ rounds, with message complexity $O(n \log n)$. If we can come up with a beeping version of this 2-ruling set algorithm, then by using the Simulation Theorem (Theorem 4 in Section 2.1) we could obtain a $k$-machine algorithm that runs in $\tilde{O}(n/k^2 + \Delta)$ rounds. By Lemma 19, $\Delta$ is bounded above by $\tilde{O}(k^{1-\epsilon})$ after Phase 1, and thus Phase 2 would run in $\tilde{O}(\frac{n}{k^2} + k^{1-\epsilon})$ rounds. The final 2-ruling set consists of the union of set $I$ obtained in Phase 1 and the 2-ruling set resulting from Phase 2.

Note that when starting to execute Phase 2, a machine $m_j$ might not be aware that some of the neighbors of one of its vertices $u$ have already been deactivated in the course of Phase 1. This does not have any effect on the round complexity of the algorithm.

This pseudocode of this two-phase algorithm is described in Algorithm 3.

---

**Algorithm 3:** TWOPHASETWORULINGSET($G = (V, E)$, $\epsilon$):

```
/* Phase 1:  Sequential Processing                                    */
```
1 $G_r \leftarrow G$;
2 $I \leftarrow \emptyset$;
3 **for** $i \leftarrow 1, \ldots, \lceil k^\epsilon \rceil$ **do**
4     Machine $m_i$ locally computes an MIS $S_i$ on its vertices;
5     $m_i$ communicates $S_i$ to all machines;
6     $S_i$ and all neighbors of nodes in $S_i$ are removed from $G_r$
7 **end**
8 $G_{low} \leftarrow$ graph induced by nodes in $G_r$ that do not have a neighbor in any machine $m_j$, $j \in [1, \lceil k^\epsilon \rceil]$;
```
/* Phase 2:  Low Message Complexity 2-Ruling Set                      */
```
9 Compute a 2-RULING SET, $I'$ for the graph $G_{low}$ using the algorithm of Pai et al. [34, 33];
10 $I \leftarrow I \cup I'$;
11 Return $I$;

---

▶ **Lemma 19.** *The time complexity of Phase 1 is $\tilde{O}(n/k^{2-\epsilon})$. Moreover, after Phase 1, the maximum vertex degree in the residual graph is $O(k^{1-\epsilon} \log n)$ with high probability.*

**Proof.** We first show the time complexity. Consider iteration $i \geq 1$ in which machine $m_i$ locally computes an MIS. By the description of the algorithm, $m_i$ sends its MIS-vertices in batches of size $O(k)$ and each machine simply broadcasts the message that it receives from $m_i$. Thus processing a single batch takes 2 rounds and since each batch processes $\Theta(k)$ vertices, we can complete iteration $i$ in $\tilde{O}(n/k^2)$ rounds. The time complexity bound follows since there are $\lceil k^\epsilon \rceil$ iterations.

We next show that only vertices that have degree $O(k^{1-\epsilon} \log n)$ remain active after Phase 1. Assume in contradiction that there is some vertex $u$ that has degree $\Omega(k^{1-\epsilon} \log n)$. Since the neighbors of $u$ were assigned to machines using random vertex partitioning, the probability that none of them is on any of the $\lceil k^\epsilon \rceil$ machines that locally processed their vertices in Phase 1 is at most

$$\left(1 - \frac{1}{k^{1-\epsilon}}\right)^{\Omega(k^{1-\epsilon} \log n)} \leq \frac{1}{n^{\Omega(1)}}.$$

Since the machine hosting $u$ knows which machines have neighbors of $u$, it will deactivate $u$ with high probability. The lemma follows by taking a union bound over all vertices. ◀

We next show that the vertices that we added when computing the local MISs in Phase 1 (and the vertices that we deactivated in the process) form a valid 2-ruling set on the induced subgraph.

▶ **Lemma 20.** *After Phase 1, the deactivated vertices form an independent set $I$ and each deactivated vertex has distance at most $2$ to some node in $I$.*

**Proof.** Clearly, no two vertices in $I$ are neighbors since all machines deactivate neighbors of nodes that were added to the (local) MISs before proceeding to the next iteration. To see that each deactivated vertex $v$ has distance at most 2, we distinguish two cases based on how $v$ was deactivated. The first possibility is that $v$ is a neighbor of some node in $I$ and we deactivated it during some iteration, in which case the distance property trivially holds. The other possibility is that $v$ was deactivated because it had a neighbor $w$ on some machine $m_j$, for $j \in [1, \lceil k^\epsilon \rceil]$. Since $m_j$ (locally) computed an MIS on its vertices, it follows that $w$ has distance at most 1 to some node in $I$ and the result follows. ◀

In Phase 2, we use the algorithm of Pai et al. [34, 33] to compute a 2-RULING SET for the graph induced by nodes that are still active and have degree at most $c \cdot \frac{n}{k^{1+\epsilon}}$ in $G$. In this algorithm, CATEGORY-1 refers to nodes that have joined the ruling set, CATEGORY-2 refers to their neighbors, and CATEGORY-3 refers nodes that have a CATEGORY-2, but not a CATEGORY-1 node in their neighborhood. This algorithm is similar to Luby's MIS algorithm, with two key differences to keep the message complexity low, at the cost of round complexity. First, the probability of a vertex $v$ being marked stays *fixed* at $1/2d(v)$ and does not increase as its neighborhood shrinks. Second, a node that is marked first uses a few messsages to determine if it should deactivate itself because it has a CATEGORY-2 node in its neighborhood. This *Checking Sampling Step* plays a key role in reducing the message complexity of this algorithm to $O(n \log n)$. Below we show that this algorithm can be implemented in the $k$-machine model in $\tilde{O}(n/k^2 + \Delta)$ rounds.

▶ **Lemma 21.** *The message-efficient 2-ruling set algorithm of [34, 33] can be implemented in the $k$-machine model in $\tilde{O}(n/k^2 + \Delta)$ rounds.*

**Proof.** The 2-ruling set algorithm of [34, 33] is not a beeping model algorithm and we cannot apply the Simulation Theorem (Theorem 1) directly to obtain an efficient $k$-machine model simulation. The difficulty is caused by steps in which a node $v$ needs to determine if a neighbor of same or higher degree has beeped. However, even in this case a machine $M$ can aggregate all the messages going to a node $v$ in a machine $M'$ by simply sending only the message to $v$ from the highest degree node it hosts. As in Theorem 1, this leads to a simulation in the $k$-machine model that runs in $\tilde{O}(msg/k^2 + T)$ rounds, where $msg$ is the message complexity and $T$ is the round complexity of the algorithm. Since Pai et al. [34, 33] have shown that $msg$ is $O(n \log n)$ and $T$ is $O(\Delta \log n)$, the result follows. ◀

▶ **Corollary 22.** *Phase 2 of Algorithm TWOPHASETWORULINGSET can be completed in $\tilde{O}(n/k^2 + k^{1-\epsilon})$ rounds.*

Combining Lemmas 19, 20, and Corollary 22, we obtain an overall running time of $\tilde{O}\left(\frac{n}{k^{2-\epsilon}} + \frac{n}{k^2} + k^{1-\epsilon}\right) = O\left(\frac{n}{k^{2-\epsilon}} + k^{1-\epsilon}\right)$, which proves the following theorem:

▶ **Theorem 23.** *For any $\epsilon$, $0 \leq \epsilon \leq 1$, a 2-ruling set can be computed in $\tilde{O}\left(\frac{n}{k^{2-\epsilon}} + k^{1-\epsilon}\right)$ rounds in the $k$-machine model.*

## 5 Future Work

Our results point to several natural followup questions:

1. Can we reconcile the gap between the $\tilde{O}(m/k^2)$ upper bound and $\tilde{\Omega}(n/k^2)$ lower bound for MIS? This seems related to the more fundamental question of showing tight bounds on the message complexity of MIS in the CONGEST KT1 model.
2. Is there an $\tilde{\Omega}(n/k^2)$ round lower bound for 2-ruling sets in the $k$-machine model? As pointed out earlier, our approach for the MIS lower bound that first proves an $\Omega(1)$-bit 2-party lower bound will not work for 2-ruling sets. Could an approach involving an $O(1)$-sized gadget distributed among 3 parties yield a lower bound for 2-ruling sets?
3. Can we improve the 2-ruling set upper bound to $\tilde{O}(n/k^2)$?
4. Can we prove non-trivial lower bounds on the space complexity of $\beta$-ruling sets in the one-pass edge-streaming model?

## A  Proofs from Section 2.2

▷ **Lemma 6** (restated). *Let $I_A$ resp. $I_B$ denote the vertices in the MIS output by Alice resp. Bob. Either $\mathbf{I}[X : I_B \mid Y] = \Omega(1)$ or $\mathbf{I}[Y : I_A \mid X] = \Omega(1)$.*

**Proof.** WLOG assume that the two nodes on Alice's side that are connected by the special edge are $u_1$ and $u_2$. Suppose Alice and Bob execute a 2-party protocol for solving MIS. Then, we have the following two cases for the nodes $\{u_3, \ldots, u_7\}$.

- Case 1: At most three nodes in $\{u_3, \ldots, u_7\}$ are in the independent set $I$. By the construction of $g$, this means that at least two nodes in $\{u_3, \ldots, u_7\}$ are not in $I$. WLOG assume that these nodes are $u_3$ and $u_4$. Now, $u_3$ and $u_4$ are not covered by any nodes on Alice's side and must be covered by their neighbors on Bob's side, i.e., $v_3$ and $v_4$ are both in $I$. Thus, $(v_3, v_4)$ cannot be a special edge on Bob's side and Alice has removed 1 of the 21 possibilities for $Y$, i.e., $H[Y|X, I_A] \leq \log_2 20$. Since $X$ and $Y$ are chosen independently, $H[X|Y] = H[X]$ and $H[Y|X] = H[Y]$. Therefore,

$$\mathbf{I}[Y : I_A | X] = H[Y|X] - H[Y|X, I_A] \geq \log_2 21 - \log_2 20 = \Omega(1).$$

- Case 2: At least 4 nodes in $\{u_3, \ldots, u_7\}$ are added to the independent set $I$. WLOG assume that these nodes are $\{u_3, u_4, u_5, u_6\}$. Then, we will look at this case from Bob's perspective. The nodes $\{v_3, v_4, v_5, v_6\}$ are all not in $I$, and at most two of these nodes are a part of the special edge. Therefore, at least two nodes in $\{v_3, v_4, v_5, v_6\}$ are not involved in the special edge on Bob's side. WLOG let $v_5$ and $v_6$ be two such vertices. Then, $v_5$ and $v_6$ are both not in $I$ and must be covered by nodes on Alice's side. Therefore, Bob knows that $u_5$ and $u_6$ must both be in $I$ and $(u_5, u_6)$ cannot be a special edge. Thus, Bob has removed 1 out of the 21 possibilities for $X$ i.e., $H[X|Y, I_B] \leq \log_2 20$ and $\mathbf{I}[X : I_B|Y] = H[X|Y] - H[X|Y, I_B] \geq \log_2 21 - \log_2 20 = \Omega(1)$.

This completes the proof of Lemma 6 and consequently Theorem 5. ◀

▷ **Lemma 8** (restated). *Either $\mathbf{I}[X : I_B \mid Y] = \Omega(n)$ or $\mathbf{I}[Y : I_A \mid X] = \Omega(n)$.*

**Proof.** Let $X(g)$ and $Y(g)$ denote the bits of $X$ and $Y$ respectively, corresponding to gadget $g$. Then, the vectors $X(g_1), \ldots, X(g_{n/14}), Y(g_1), \ldots, Y(g_{n/14})$ are all mutually independent. Using the additive property of mutual information under independence, we get that

$$\mathbf{I}[X : I_B \mid Y] = \sum_{i=1}^{n/14} \mathbf{I}[X(g_i) : I_B \mid Y]. \tag{2}$$

Since $X$ and $Y$ are independent, $H[X(g_i) \mid Y] = H[X(g_i)] \geq H[X(g_i) \mid Y(g_i)]$. Moreover, since $Y(g_i)$ and $I_B(g_i)$ are subsets of $Y$ and $I_B$ respectively, we get

$$H[X(g_i)|Y, I_B] \leq H[X(g_i)|Y(g_i), I_B(g_i)]$$

Combining this with the definition of mutual information, we obtain

$$\begin{aligned}\mathbf{I}[X(g_i) : I_B|Y] &= H[X(g_i)|Y] - H[X(g_i)|I_B, Y] \\ &\geq H[X(g_i)|Y(g_i)] - H[X(g_i)|Y(g_i), I_B(g_i)] \\ &= \mathbf{I}[X(g_i) : I_B(g_i)|Y(g_i)].\end{aligned}$$

Using this result in Equation (2), we get

$$\mathbf{I}[X : I_B|Y] \geq \sum_{i=1}^{\frac{n}{14}} \mathbf{I}[X(g_i) : I_B(g_i)|Y(g_i)],$$

and, similarly,

$$\mathbf{I}[Y : I_A | X] \geq \sum_{i=1}^{\frac{n}{14}} \mathbf{I}[Y(g_i) : I_A(g_i) | X(g_i)].$$

From Theorem 6, we know that for each gadget $g_i$, either $\mathbf{I}[Y(g_i) : I_A(g_i)|X(g_i)] = \Omega(1)$ or $\mathbf{I}[X(g_i) : I_B(g_i)|Y(g_i)] = \Omega(1)$. Thus, either $\mathbf{I}[X : I_B|Y] = \Omega(n)$ or $\Omega(n)$.　◀

▷ **Theorem 9 (restated).** For a constant $\varepsilon > 0$, any $\varepsilon$-error (possibly randomized) MIS algorithm in the $k$-machine model has round complexity at least $\tilde{\Omega}(\frac{n}{k^2})$.

**Proof.** Consider an MIS algorithm, $ALG$, in the $k$-machine model that runs in $\tilde{o}(\frac{n}{k^2})$ rounds with error probability at most $\varepsilon$. We use a simulation argument similar to the one used in the proof of Theorem 4.2 of [36]. That is, we ensure that Alice and Bob can use their inputs to simulate $ALG$ on $G_L$. More specifically, if we partition the $k$ machines into two sets $P_A := \{m_1, m_3, \ldots, m_{k-1}\}$ and $P_B := \{m_2, m_4, \ldots, m_k\}$, then Alice and Bob can simulate $ALG$ for machines in $P_A$ and $P_B$ respectively. However, notice that our current setup requires all the $u$ nodes to be hosted at Alice and all the $v$ nodes to be hosted at Bob i.e., all the $u$ nodes must be assigned to machines in $P_A$ and all the $v$ nodes must be assigned to machines in $P_B$. Such partitions form less than a $\frac{1}{2^{k/2}}$ fraction of all graph partitions. Since $ALG$ is an $\varepsilon$ error algorithm, it may not output an MIS in all such partitions that assign $u$ nodes to machines in $P_A$ and $v$ nodes to machines in $P_B$. In other words, we need a communication complexity model that takes into account such issues. Therefore, similarly to [26], we assume the random partition 2-party communication model.

In the random partition two party communication model, instead of all the digits in $X$ going to Alice and all the digits in $Y$ going to Bob, each digit of a single input $Z$ is assigned randomly between Alice and Bob. For our purposes, we let $Z := (X, Y)$, where both $X$ and $Y$ are valid $n/2$ digit vectors. Thus, under the random partition two party communication model, each digit of $X$ and $Y$ is assigned to Alice or Bob with probability $1/2$. If Alice gets digit $x_i$, then Alice will use shared randomness to assign $u_i$ to a random machine, $m_j$ in $P_A$. If Alice gets $y_i$ also, then Alice will know which machine $v_i$ was assigned to and can tell $m_j$ about the machine hosting $v_i$. On the other hand, if Bob gets digit $y_i$ then Alice still knows which machine $v_i$ was assigned to since Alice and Bob use shared randomness to make these decisions. Finally, since Alice knows the value of digit $x_i$, she knows whether $u_i$ has a neighbor in addition to $v_i$, and the machine hosting this neighbors. Thus, for all the nodes that Alice simulates, she knows their neighbors and the machines hosting these neighbors. Similarly, for each node that Bob hosts, he knows all of the node's neighbors and the machines hosting these neighbors. Notice that in this process, the probability that a node $u$ ends at a machine $m$ is $1/k$ since $u$ is assigned to Alice or Bob with probability $1/2$, and subsequently Alice or Bob will assign $u$ to a specific machine $m$ with probability $2/k$. Thus, this process gives us a random partition of the $n$ nodes over the $k$ machines, and Alice and Bob can simulate $ALG$.

Under the random partition of $X$ and $Y$, let $W_A$ and $W_B$ denote the digits received by Alice and Bob respectively. Since $ALG$ computes an MIS of $G$, our goal is to bound the mutual information between $W_A$ and $W_B$ and any MIS of $G_L$. Consider a single gadget, $g_i$ with value $(x_{7i+1}, y_{7i+1})$. Notice that if both Alice and Bob get at least a single digit each from $(x_{7i+1}, \ldots, x_{7i+7})$ and $(y_{7i+1}, \ldots, y_{7i+7})$, then they both know the complete topology of gadget $g_i$. The probability that this bad event doesn't happen is $2^{-13}$. We call such a gadget *good*, and define *Good* to be the *set of all good gadgets*. The expected number of good gadgets in a node partition is $2^{-13} \cdot n/14$. Since the events that gadgets are good are independent, we can use a standard Chernoff bound to show that the number of good gadgets is $\Omega(n)$ whp i.e., at least a $(1 - o(1/n))$ fraction of the graph partitions have $\Omega(n)$ *good* gadgets. Assuming that $ALG$ fails to compute an MIS for at most a $\varepsilon$ fraction of these, it must succeeds in doing so for at least a $(1 - o(1/n) - \varepsilon)$ fraction of graph partitions.

We are now ready to compute a lower bound the mutual information between any MIS of $G_L$ and the inputs $W_A$ and $W_B$. Let $I$ denote an MIS of $G_L$ and let $I_A$ and $I_B$ denote subsets of $I$ hosted at Alice and Bob respectively.

Then, the mutual information between Bob's input, $W_B$, and the MIS conditioned on Alice's input, $W_A$, is

$$\mathbf{I}[W_B : I_A | W_A] = H[W_B | W_A] - H[W_B | I_A, W_A].$$

Let $W_B(g)$ and $W_A(g)$ denote the digits in $W_B$ and $W_A$ that correspond to gadget $g$. Then, the inputs $W_A(g_i), \ldots, W_A(g_{\frac{n}{14}})$ are all mutually independent. Using the additive property of mutual information under independence,

$$\mathbf{I}[W_A : I_B | W_B] = \sum_{i=1}^{\frac{n}{14}} \mathbf{I}[W_A(g_i) : I_B | W_B]$$
$$\geq \sum_{g \in Good} \mathbf{I}[W_A(g) : I_B | W_B].$$

Notice that for *good* gadgets, we have the same situation as in the proof of Lemma 8. Using this result, we get that

$$\mathbf{I}[W_A : I_B \mid W_B] = \sum_{g \in Good} \mathbf{I}[W_A(g) : I_B(g) \mid W_B(g)]$$

and

$$\mathbf{I}[W_B : I_A \mid W_A] = \sum_{g \in Good} \mathbf{I}[W_B(g) : I_A(g) \mid W_A(g)]$$

We conclude the proof of the theorem by using the fact that the number of *good* gadgets is $\Omega(n)$ *whp*, and that Theorem 5 implies that, for each *good* gadget, either $\mathbf{I}[W_A(g) : I_B(g) | W_B(g)]$ or $\mathbf{I}[W_B(g) : I_A(g) | W_A(g)]$ is $\Omega(1)$. ◀

## B Proofs from Section 3.1

▷ **Lemma 10 (restated).** For $1 \leq i \leq \beta$, the maximum degree of nodes in $P_i$ is at most $\Delta^{1-(i-1)/\beta}$.

**Proof.** The claim is trivially true for $i = 1$. For $i > 1$, every node in $P_{i-1}$ that has degree greater than $\Delta^{1-(i-1)/\beta}$ is placed in $T_i$ and removed from $P_i$ (Steps (7-8)). ◀

▷ **Lemma 11.** For $2 \leq i \leq \beta$, the maximum degree of the induced graph $G[M_i]$ is at most $\tilde{O}(\Delta^{1/\beta})$ whp.

**Proof.** Consider an arbitrary $i$, $2 \leq i \leq \beta$. By Lemma 10, a node in $P_{i-1}$ has degree at most $\Delta^{1-(i-2)/\beta}$. Nodes in $P_{i-1}$ independently join $M_i$ with probability $\Theta(\log n/\Delta^{1-(i-1)/\beta})$. Therefore, using Chernoff bounds, we see that $G[M_i]$ has maximum degree at most $\tilde{O}(\Delta^{1/\beta})$ whp. ◀

▷ **Lemma 12 (restated).** The communication in Steps (5) and (10) can each be completed in $\tilde{O}(n\Delta^{1/\beta}/k^2)$ rounds whp. The MIS computation in Steps (6) and (11) can each be completed in $\tilde{O}(n\Delta^{1/\beta}/k^2)$ rounds whp.

**Proof.** We first focus on Step (5). Since nodes are distributed uniformly at random among the $k$ machines and nodes in $P_{i-1}$ are independently marked with probability $\Theta(\log n/\Delta^{1-(i-1)/\beta})$, using Chernoff bounds we see that $\tilde{O}(\frac{n}{k \cdot \Delta^{1-(i-1)/\beta}})$ nodes are marked at each machine. By Lemma 10, nodes in $P_{i-1}$ have maximum degree at most $\Delta^{1-(i-2)/\beta}$. This means that the volume of messages to be sent at each machine is $\tilde{O}(n \cdot \Delta^{1/\beta}/k)$ whp. Similarly, the volume of messages to be received at each machine is $\tilde{O}(n \cdot \Delta^{1/\beta}/k)$ whp. Using a randomized routing algorithm (Algorithm 1) we can deliver all of these messages in $\tilde{O}(n \cdot \Delta^{1/\beta}/k^2)$ whp.

We now focus on Step (10). Each machine hosts $\tilde{O}(n/k)$ nodes whp and by Lemma 10, nodes in $P_\beta$ have degree at most $\Delta^{1/\beta}$. Therefore, the volume of messages to be sent and to be received by a machine is $\tilde{O}(n\Delta^{1/\beta}/k)$ whp. Using a randomized routing algorithm (Algorithm 1) we can deliver all of these messages in $\tilde{O}(n \cdot \Delta^{1/\beta}/k^2)$ rounds whp.

By Lemmas 10 and 11, the MIS computation in Steps (6) and (11) is on induced graphs with maximum degree at most $\tilde{O}(\Delta^{1/\beta})$. Also, due to communication in previous steps, each node participating in an MIS computation knows which neighbors are also participating in the MIS computation. Therefore, we can simply use the $k$-machine MIS algorithm (see Theorem 4) that runs in $\tilde{O}(m/k^2)$ on an $m$-edge graph to complete the MIS computations in $\tilde{O}(n\Delta^{1/\beta}/k^2)$ rounds whp. ◀

▷ **Lemma 13 (restated).** Every node in $V$ is at most $t$ hops from some node in $(\cup_{j=2}^{\beta} I_j) \cup I$.

**Proof.** The set $V$ of nodes of the input graph can be written as $(\cup_{i=2}^{t}(M_i \cup T_i)) \cup P_t$. Each node in $M_i$ is at most one hop away from a node in $I_i$ and similarly each node in $P_t$ is at most one hop away from a node in $I$.

We now show that each node in $T_i$ is at most $i$ hops away from some node in $\cup_{j=1}^{i} I_j$. If an unmarked node $v$ joins $T_i$ because it has a neighbor in $M_i$, then it is at most 2 hops away from some node in $I_i$. Otherwise, an unmarked node $v$ joins $T_i$ because it has degree greater than $\Delta^{1-(i-1)/t}$. If all neighbors of $v$ participate in the sampling step in Iteration $i$, then $v$ is guaranteed (whp) to have a neighbor in $M_i$. But, we know this not to be the case. Therefore, it must be the case that $v$ has at least one neighbor that was deactivated in a previous iteration $j < i$, i.e., joined $T_j$. Inductively assuming that a node deactivated in Iteration $j$ is at most $j$ hops from $I_1 \cup I_2 \cup \cdots \cup I_j$, we see that node $v$ is at most $j + 1 \leq i$ hops away from some node in $\cup_{j=1}^{i} I_j$. ◀

## C  Proofs from Section 3.2

▷ **Lemma 16 (restated).** The algorithm uses $O\left(\beta \cdot n^{1+1/2^{\beta-1}} \log n\right)$ space with high probability.

**Proof.** We first restrict our attention to nodes that were *not* moved during the post-processing step. We use a simple charging argument to analyze the space usage of the algorithm. If some node $u$ is not yet covered and we observe an incident edge $(u, v)$ in the stream, then we *charge edge $(u, v)$ to $u$*. Note that if $v$ is not yet covered at this point, we also charge $(u, v)$ to $v$. Since the algorithm only stores edges for nodes that are not yet covered, it will be sufficient if we obtain an upper bound on the number of charged edges to bound the space usage.

For each $u \in P_1$, we charge up to $\mu_1 = O(n^{q_1} \log n)$ edges, and hence the space required for edges in $P_1$ is at most $O\left(n^{1+q_1}\right) = O\left(n^{1+1/2^{\beta-1}}\right)$. Now suppose that $u \in P_i$, for some

$i \in [2, \beta]$. By the description of the algorithm,

$$\mathbf{Pr}[\ell(u) \geq i\,] \leq \prod_{j=1}^{i-1} \frac{1}{n^{1/2^{\beta-j}}}, \tag{3}$$

and hence $\mathbf{E}\left[|P_i|\right] \leq n \prod_{j=1}^{i-1} \frac{1}{n^{1/2^{\beta-j}}}$. Since we sample the levels for each vertex independently, we can use a standard Chernoff bound to show that, whp,

$$\begin{aligned}
|P_i| &= O\left(n \log n \cdot \prod_{j=1}^{i-1} \frac{1}{n^{1/2^{\beta-j}}}\right) \\
&= O\left(n^{\left(1 - \frac{1}{2^{\beta-1}} \sum_{j=1}^{i-1} 2^{j-1}\right)} \log n\right) \\
&= O\left(n^{\left(1 - \frac{1}{2^{\beta-1}}(2^{i-1} - 1)\right)} \log n\right) \\
&= O\left(n^{1 - q_i + 1/2^{\beta-1}} \log n\right).
\end{aligned} \tag{4}$$

In the remainder of the proof, we use (4) to bound the number of edges that were charged. Suppose that $i \in [2, \beta - 1]$, i.e., $u \notin P_\beta$, then, given that we charge $\mu_i$ edges to each node in $P_i$ ($i < \beta$), it follows by (4) that we store $O(n^{1+1/2^{\beta-1}} \log n)$ edges in total for all nodes in $P_i$ with high probability. Next, consider the case $u \in P_\beta$, i.e., $i = \beta$ and recall that each node in $P_\beta$ is charged for all its incident edges. Recalling $q_\beta = 1$, (4) yields $|P_\beta| \leq O\left(n^{1/2^{\beta-1}} \log n\right)$. It follows that we store at most $O\left(n^{1+1/2^{\beta-1}} \log n\right)$ edges for nodes in $P_\beta$ whp.

Finally, consider a node $w$ that was moved during the post-processing step. Given that the algorithm only moved $w$ because it was not covered, it follows that we charged at most $O(n^{q_i} \log n)$ edges to $w$, where $i$ is $w$'s level *before* the post-processing step. Considering that we move at most $O(|P_i|)$ vertices from $P_i$, it follows from (4) that we obtain the same space bound as for non-moved nodes. ◀

**Algorithm 4:** Algorithm `2-rulingset-msg`: code for a node $v$. $d(v)$ is the degree of $v$.

1 $\text{status}_v = \text{UNDECIDED}$;
2 **while** $status_v = \text{UNDECIDED}$ **do**
3     **if** *v receives a message from a CATEGORY-1 node* **then**
4        Set $\text{status}_v = \text{CATEGORY-2}$;
5     **end**
6     **if** *v is UNDECIDED* **then** it marks itself with probability $\frac{1}{2d(v)}$ ;
7     **if** *v is marked* **then**
8        (*Checking Sampling Step:*) Sample a set $A_v$ of $4\log(d(v))$ random neighbors independently and uniformly at random (with replacement) ;
9        Find the categories of all nodes in $A_v$ by communicating with them;
10       **if** *any node in $A_v$ is a CATEGORY-2 node* **then**
11          Set $\text{status}_v = \text{CATEGORY-3}$;
12       **end**
13       **else**
14          (*(Local) Broadcast Step:*) Send the marked status and $d(v)$ value to *all* neighbors;
15          If $v$ hears from an equal or higher degree (marked) neighbor then $v$ unmarks itself;
16          If $v$ remains marked, set $\text{status}_v = \text{CATEGORY-1}$;
17          Announce status to all neighbors;
18       **end**
19     **end**
20 **end**